\documentstyle[psbox]{elsart}

\begin{document}
\begin{frontmatter}
\title{Direct measurement of sub-pixel structure of the EPIC MOS CCD on-board the XMM/NEWTON satellite}
\date{}

\author[osaka]{J. HIRAGA\thanksref{jsps}}
\author[osaka]{H.TSUNEMI}
\author[le]{A.D.SHORT}
\author[le]{A.F.ABBEY}
\author[le]{P.J.BENNIE}
\author[le]{M.J.L.TURNER}

\thanks[jsps]{Partially supported by JSPS Research Fellowship for Young
 Scientists, Japan.}

\address[osaka]{Department of Earth and Space Science, Graduate School
of Science, Osaka University, 1-1 Machikaneyama-cho, Toyonaka, Osaka
5600043, Japan \\
CREST, Japan Science and Technology Corporation (JST)\\
}
\address[le]{Space Reserch Centre, University of Leicester, Leicester,
 LE1 7RH, UK}

\begin{abstract}

We have used a mesh experiment in order to measure the sub-pixel structure
of the EPIC MOS CCDs on-board the XMM/NEWTON satellite.  The EPIC MOS CCDs
have 40\,$\mu$m-square pixels which have an open electrode structure in
order to improve the detection efficiency for low-energy X-rays.  We
obtained restored pixel images for various X-ray event grades (e.g.
split-pixel events, single pixel events, etc.) at various X-ray
 energies.

We confirmed that the open electrode structure results in a distorted
horizontal pixel boundary.  The open electrode region generates both
single pixel events and vertically split events, but no horizontally
split events.  Because the single pixel events usually show the best
energy resolution, we discuss a method of increasing the fraction of
single pixel events from the open electrode region.  Furthermore, we
have directly measured the thickness of the electrodes and dead-layers
by comparing spectra from the open electrode region with those from the
other regions: electrodes, electrode finger and channel stop. We can say
that EPIC MOS CCDs are more radiation hard than front-illumination chips
of ACIS on-board Chandra X-ray Observatory because of their extra
absorption thickness above the charge transfer channel. We calcurated
the mean pixel response and found that our estimation has a good agreement
with that of the ground calibration of EPIC MOS CCD.

\end{abstract}


\end{frontmatter}

\section{Introduction}

Charge-coupled devices (CCDs) in use in X-ray astronomy combine
moderate energy resolution with good spatial resolution\cite{frazer89}.
Thanks to these characteristics, they have become a standard X-ray
photon counting detector. When an X-ray photon is photoabsorbed inside
a CCD, a number of electrons proportional to the incident X-ray energy
are liberated.  In this way, the energy of an X-ray photon can be
measured.  Because optical photons can produce only a few electrons,
however, CCDs have no energy resolution at optical wavelengths.

The charge cloud produced by an X-ray photon inside the CCD drifts
within the depletion region to the bottom of the potential well of the
given pixel, resulting in a detected X-ray event.  Due to the diffusion
process, the charge cloud has a finite size, which can result in the
event being detected in more than one pixel. X-ray events may be
therefore classified by \lq grade\rq\ according to the number of
pixels in which they are detected.  When the entire charge cloud is
collected within one pixel, for example, it is referred to as a \lq
single pixel event\rq. When the cloud splits into an adjacent horizontal
pixel, it is referred to as a \lq horizontally split event\rq.
 Similarly, when the cloud splits into an adjacent vertical pixel, it is
referred to as a \lq vertically split event\rq.

A CCD consists of a two dimensional array of small pixels. The spatial
resolution is limited by the pixel size which is typically several tens
of $\mu$m.  The electrode structure, which comprises layers of
poly-silicon and silicon-oxide, results in a non-uniformity of detection
efficiency over the pixel.  In order to measure the structure of one
pixel directly, it is therefore necessary to determine the X-ray
interaction position on a scale smaller than the pixel size. Recently,
we have developed a new technique which allows us to specify the X-ray
interaction position with sub-pixel resolution using a two-dimensional
mesh containing small holes (much smaller than the CCD pixel size) which
are periodically spaced~\cite{tsunemi97} .  The sub-pixel structure of
various types of CCD (ASCA SIS~\cite{yoshita98}, CHANDRA
ACIS~\cite{pivovaroff98} and HPK CCD~\cite{tsunemi98}) have been
measured using this method.

The X-ray response of a CCD is very sensitive to the thickness of the
gate structures within the pixel and is also non-uniform within the
pixel. The time-averaged output therefore represents a mean pixel
response, rather than the response at any given location within the
pixel. Thus the CCD response consists of various parameters which are
difficult to measure separately.  We have therefore developed the mesh
technique in order to measure the response (pixel structure) of CCDs
with sub-pixel resolution.  This technique has been previously
applied to the CCDs \cite{alex98} employed in the XMM/NEWTON observatory
 by Tsunemi et al\cite{turner98}.  In this paper, we report on
measurements of the sub-pixel structure of the EPIC MOS CCD with
substantially improved spatial resolution.

\section{The EPIC MOS CCD}

The XMM/NEWTON satellite was developed by the European Space Agency
(ESA) which comprises a membership of 13 European countries.  It was
launched into a relatively high-earth orbit in December, 1999.  Among
its instruments are the 3 EPIC imaging spectrometers, which reside in
the focal plane, at the foci of the three mirror modules.  All carry
silicon CCD detectors.  One of the cameras utilizes back-illuminated PN
CCDs with 150\,$\mu$m square pixels and was developed at the Max Plank
Institute~\cite{elmar99}.  The other two cameras carry MOS CCDs\,\,(EEV
CCD 22s) which were developed primarily by the X-ray Astronomy Group at
Leicester University and Marconi Applied Technologies (formerly EEV) in
the United Kingdom~\cite{alex98}.  The EPIC MOS CCD is a frame transfer,
front-illuminated device.

\begin{figure}[n]
\begin{center}
\psbox[r,xsize=0.5#1]{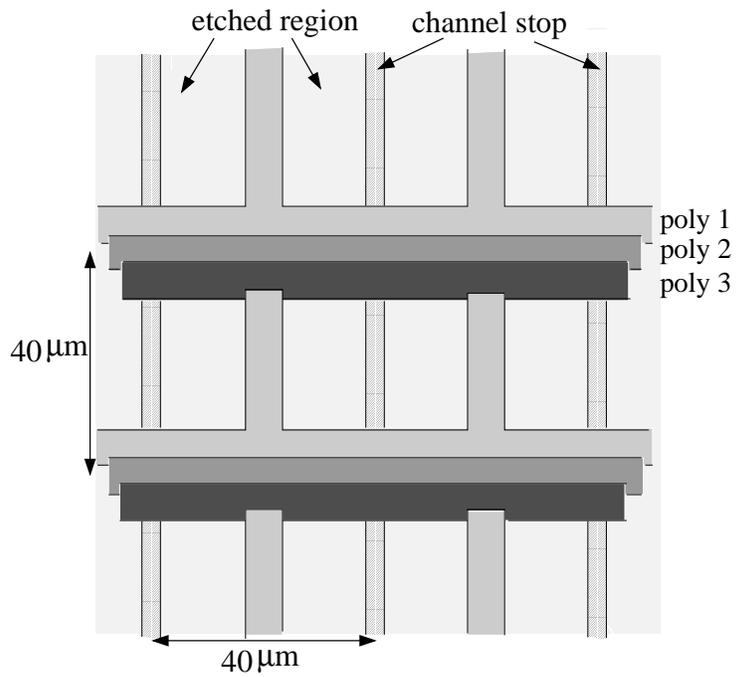}
\caption{Schematic structure inside the pixel of an EPIC MOS CCD. There
 are three electrodes: one is partly etched in order to improve
 detection efficiency at low energy.}
\label{1pix_struct} \hspace{-8mm}
\end{center}
\end{figure}

The EPIC MOS CCD is a three phase device.  The electrodes, or gates
(poly1, poly2 and poly3) are shown schematically in
Figure\ref{1pix_struct}. The thickness of depletion region is approximately
37\,$\mu$m with nominal clock voltages and substrate bias \cite{alex98}.

The most important feature of the EPIC MOS CCD is its \lq open electrode
structure\rq.  In order to improve the detection efficiency at low
energies, one of the gates, poly3, has been enlarged by partially
removing the front face, leaving two \lq holes\rq\ in each pixel.  These
holes cover 40\,\% of the pixel area and are separated by a central
electrode \lq finger\rq\ comprising polysilicon, oxide and nitride
layers.  A P-plus dopant is implanted in the etched areas, which pins
the surface potential to the substrate potential.
 
\section{Experiment}

\subsection{Experimental Setup}

\begin{figure}[htbn]
\begin{center}
\psbox[r,xsize=0.35#1]{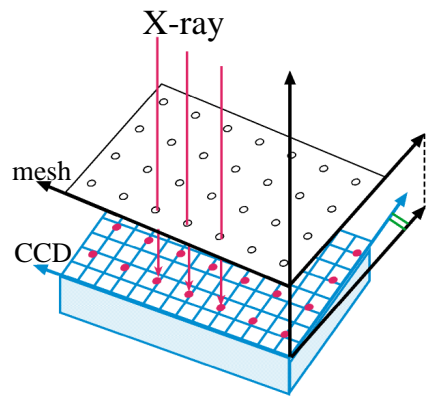}
\caption{Schematic view of the mesh experiment showing the orientation
 of the mesh with respect to the CCD. The X-ray landing position is
 restricted by the mesh hole.}
\label{consept} \hspace{-8mm}
\end{center}
\end{figure}

A detailed explanation of our mesh experiments may be found in the
literature \cite{tsunemi97}\cite{tsunemi98}.  Figure~\ref{consept} gives a
schematic view of the mesh experiment.  The mesh experiment consists of a
CCD, a metal mesh and a pseudo parallel X-ray beam.  The metal mesh
employed has periodically spaced holes which are much smaller than the
pixel size.  The hole spacing is an integer multiple of the CCD pixel
size.  The mesh is placed just above the CCD surface, as close to the CCD
as practically possible.

The mesh must have an orientation which is slightly rotated with respect
to the CCD so that the shadow of the mesh hole on the CCD gradually
shifts its position inside the pixel as shown in Fig. \ref{consept}.
In this way, over the CCD dimensions, the X-rays passing through the
mesh holes periodically sample the entire pixel. This produces a moire
pattern from which the relative alignment between the mesh and the CCD
may be determined~\cite{yoshita98}.  An X-ray event detected by the CCD
must have come through one of the mesh holes. Taking into account the
hole spacing and the pixel size, we can unequivocally determine the hole
location for individual X-ray events.  We can therefore calculate the
X-ray interaction position within the CCD with sub-pixel resolution.
The accuracy is limited by the effective mesh hole size which is
slightly bigger than the geometrical shape of the hole due to
diffraction.

\begin{figure}[htbn]
\begin{center}
\psbox[r,xsize=0.35#1]{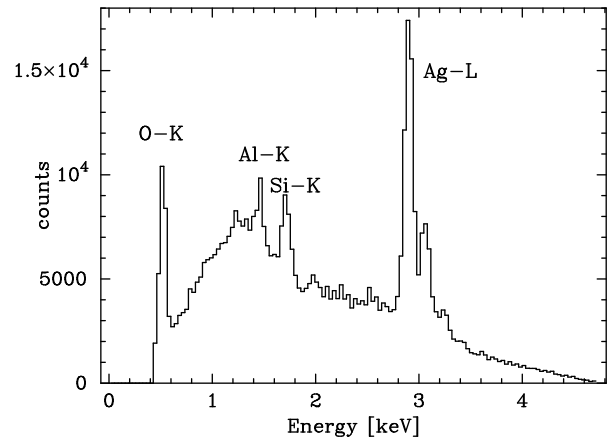} \caption{Energy spectrum
obtained using the whole CCD data. There are several characteristic
emission lines, O-K (0.52\,keV), Al-K (1.5\,keV), Si-K (1.8\,keV) and
Ag-L(2.9\,keV), superposed on a continuum extending up to 5\,keV. } 
\label{spec_all} \hspace{-8mm}
\end{center}
\end{figure}

The experiment was performed in a CCD test facility at Leicester
University.  The EEV CCD 22 has 600\,$\times$\,600 pixels with each
pixel being 40\,$\mu$m square.  A gold mesh was employed which has a
thickness of 10\,$\mu$m and small holes of 2\,$\mu$m diameter. The
spacing between the holes was 120\,$\mu$m; just three times the pixel
size.  We placed the mesh about 0.5\,mm above the CCD surface and
rotated it by about 1.$\!\!^{\circ}$7 from the CCD.  The X-ray generator
manufactured by KEVEX Inc. was approximately 3\,m from the CCD and
several fluorescence targets were used, generating characteristic X-rays
as well as a Bremsstrahlung spectrum.  Figure\,\ref{spec_all} gives an
example the spectrum obtained by the mesh experiment using a Ag target
with a voltage of 5\,kV.  There are several characteristic emission
lines, O-K (0.52\,keV), Al-K (1.5\,keV), Si-K (1.8\,keV) and
Ag-L(2.9\,keV), superposed on a continuum extending up to 5\,keV.

The CCD operating conditions were almost identical to those employed on
the XMM/NEWTON satellite.  During frame integration we apply 8\,V (high
voltage) to poly3 (in Fig.\ref{1pix_struct}) only.  The other two gates
remained un-biased.  The CCD chip was cooled to $-$100$^{\circ}$C using
liquid nitrogen and was driven using duplicate flight electronics.

Tsunemi et al. (1999) performed the mesh experiments using a similar
experimental setup~\cite{tsunemi99d}.  Their experiment used an existing
copper mesh with a thickness of 10\,$\mu$m and holes of 4\,$\mu$m
diameter. The spacing between the holes was 48\,$\mu$m which is not a
multiple of the pixel size.  The effective mesh hole at that time was
7\,$\mu$m in diameter.  For this experiment, we have improved the
conditions by fabricating a new mesh which may be positioned much closer
to the CCD surface with holes 2\,$\mu$m in diameter. In the new
configuration, the effective mesh hole gives about a factor of 3
improvement in spatial resolution over the previous experiment.
Furtheremore, the hole spacing is much greater, and is an integer
multiple of the pixel size, which makes the determination of the X-ray
interaction location within a given pixel much clearer.

\section{Data Analysis and Discussion}

\subsection{Image Restoration}

The mesh technique allows us to determine the X-ray interaction position
within the CCD pixel.  Furthermore, it samples the entire pixel which
enables us to restore the pixel images for the various X-ray energies as
well as the various X-ray event-types.  In the data analysis, we pick out
the characteristic X-ray energies thanks to their good statistics (as
summarized in table~\ref{eev_exp} in order of attenuation length in
SiO$_2$).  We then construct restored images for various X-ray
event-types; (a) single pixel events, (b) vertically split events, (c)
horizontally split events and (d) all X-ray events, as shown in
Figure~\ref{restored}.  In this figure, each image represents 2$\times$2
pixels of the CCD, with the dashed square corresponding to the pixel size
of 40\,$\mu$m square.  Brighter regions correspond to a higher detection
efficiency.

Looking at the pixel image restored using all X-ray events (the right hand
column), we can clearly see absorption features within the pixel.  In
particular, the two etched regions in each pixel are clearly visible in
the O-K image, since O-K has the shortest attenuation length of the X-ray
energies employed.  Moving to greater attenuation lengths, the etched
region becomes less obvious.  It is difficult to see the enhanced region
in the Ag-L image which is almost free from absorption within the gate
structure.

In the restored image using single pixel events, we clearly see the
enhanced region only in the O-K image.  Other than the gate structure,
we notice that there are three regions in each pixel: the regions
generating single pixel events, vertically split events and horizontally
split events.  Split events are generated in the regions within the
pixel where the charge cloud splits into adjacent pixels.  We can
clearly see that the horizontal pixel boundary (giving rise to
vertically split events) is not a straight line but is instead a wavy
line.  In contrast to this, the vertical pixel boundary is governed by
the channel stops, which include a P-plus implant in order to make an
electric potential \lq barrier\rq\, which results in a normal
(i.e. straight line) shape for the pixel boundary.  There are small gaps
in the region generating horizontally split events.  These correspond to
the pixel corners where the 3- or 4-pixel events are generated.

The wavy boundary is due to the fact that the poly3 gate is etched so that
it has a finger structure as shown in Fig.\ref{1pix_struct}.  In other
words, the potential generated by poly3 during integration defines a wavy
boundary between adjacent pixels rather than a straight line.  The shape
of the region generating vertically split events also depends on the
attenuation length of the X-rays indicating the depth dependence of the
electric field inside the CCD.  It can be compared with that expected from
the model calculation. 


\subsection{Spectra from Various Regions inside the Pixel}

\begin{figure}[htbn]
\begin{center}
\psbox[r,xsize=0.6#1]{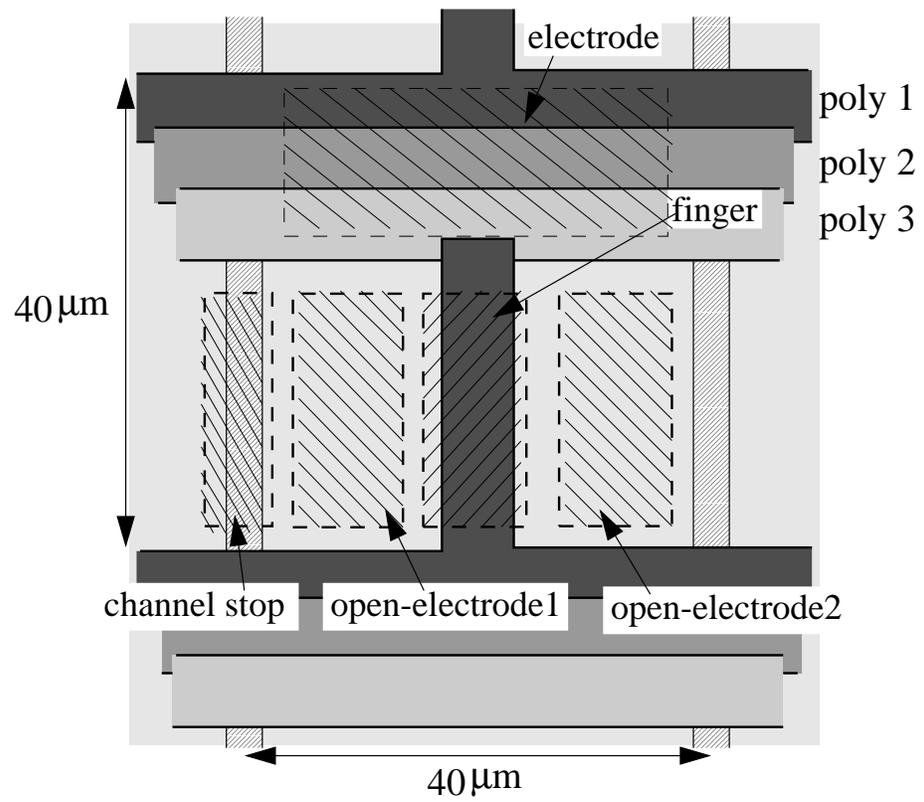}
\caption{Five selected regions within the pixel are shown.  They are taking
into account the
 electrodes or channel stop structure to compare spectra to each other.}
\label{spec_region} 
\end{center}
\end{figure}

\begin{figure}[htbn]
\begin{center}
\psbox[xsize=0.5#1]{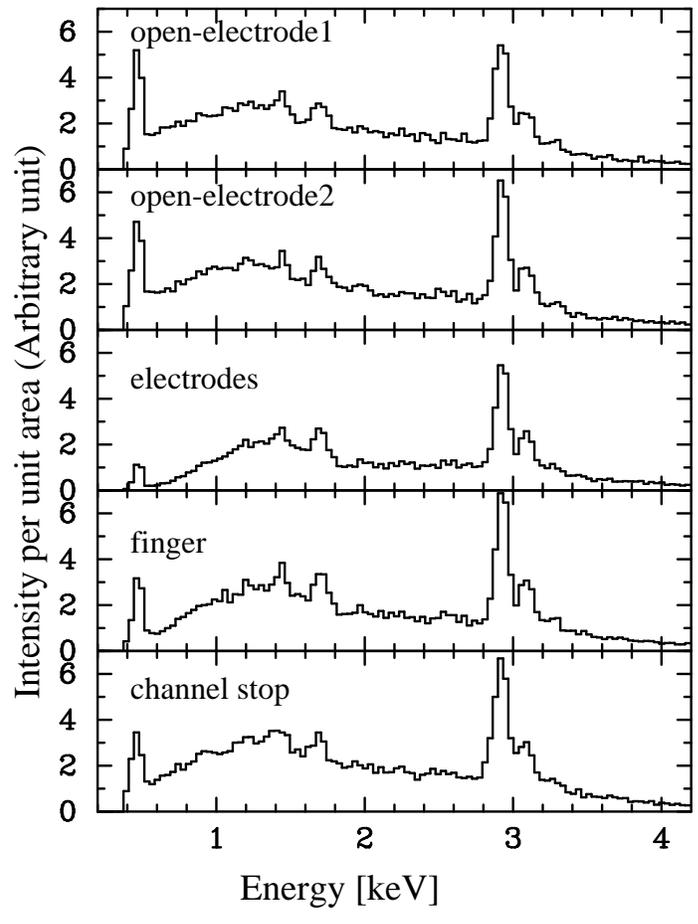}
\caption{Spectra from five selected regions within the pixel. }
\label{regspec} 
\end{center}
\end{figure}

In the mesh experiment, we can identify both the energy and the
interaction position within the pixel for an individual X-ray photon. We
can therefore extract X-ray spectra from any region within the pixel.
We selected five regions within the pixel in order to measure the pixel
structure of the CCD.  These are labeled in Figure~\ref{spec_region} as
\lq finger\rq,\, \lq channel stop\rq,\, \lq electrode\rq,\, \lq open
electrode 1\rq\, and \lq open electrode 2\rq,\, respectively.  Among
them, the channel stop and the finger are so narrow that the selected
regions partially overlap with the open electrode region due to the
finite spatial resolution of the experiment.  Figure~\ref{regspec} shows
the spectra for each region using the same data in Fig.~\ref{spec_all}.

\begin{figure}[htbn]
\begin{center}
\psbox[r,xsize=0.3#1]{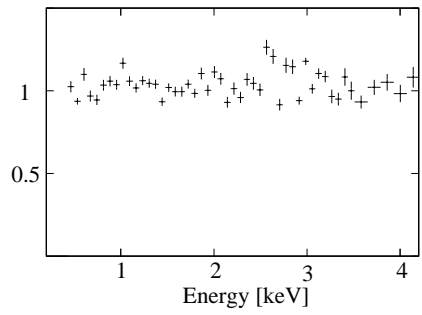}
\caption{Ratio between the spectrum obtained at the open electrode 1 and
that obtained at the open electrode 2. }
\label{open} 
\end{center}
\end{figure}

\begin{figure}[htbn]
\begin{center}
\psbox[r,xsize=0.3#1]{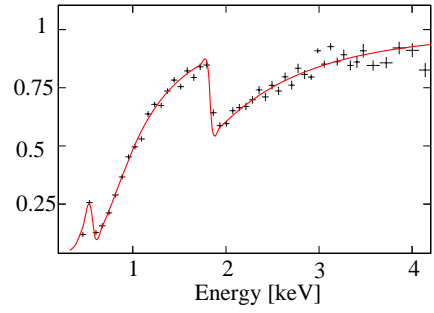}
\psbox[r,xsize=0.3#1]{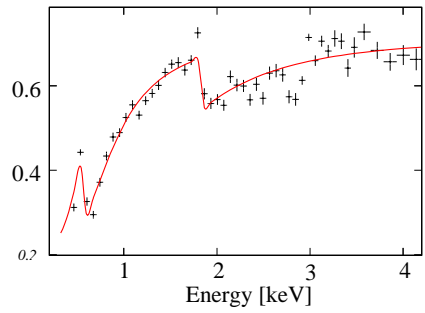}
\psbox[r,xsize=0.3#1]{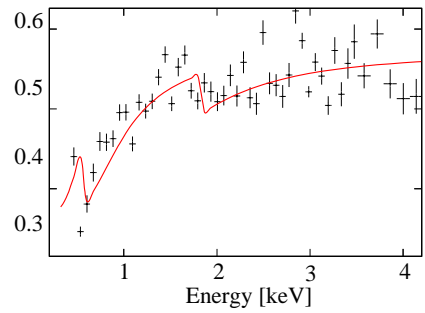}
\caption{Ratios between the spectrum of the open electrodes and those of
selected regions:\,upper, middle and lower panels represent results of
 electrodes, finger and channel stop, respectively. Solid lines represent the best fit absorption model.}
\label{trans} 
\end{center}
\end{figure}

We first compared spectra for the open electrode 1 and 2.  We confirmed
that the ratio shown in Figure~\ref{open} is almost constant and
independent of energy.  Because these two regions must show the highest
efficiency at low energies, we added them together to generate a
standard spectrum for the open electrode with which we compare other
data.  Figure~\ref{trans} shows the ratio between the spectra for the
open electrode and those for other three regions:\,upper panel, middle
panel and lower panel are results of electrodes, finger and channel
stop, respectively. Since these figures represent the extra absorption
features in each region, we fitted them with a relatively simple
absorption model. The model contains a normalization, a constant
component and absorption features for Si and SiO$_2$.  The normalization
is required due to the difference in area of each region and the
constant comes from the area of overlap with the open electrode region.
In this way, we obtained the thickness of Si and SiO$_2$ for each
region.  The best fit results are shown in Fig.~\ref{trans} by solid
lines and summarized in Table~\ref{result}.

\subsection{Radiation Hardness}
In June 1999 the Chandra X-ray Observatory (CXO) was launched into a a
high-earth orbit, similar to that of the XMM/NEWTON satellite.  It is
equipped with a CCD camera (ACIS) consisting of ten CCD chips: eight
front-illuminated\,(FI) CCDs and two back-illuminated\,(BI) CCDs.  In august, it was reported that a substantial degradation of the
CCD had occurred~\cite{odel00}.  The degree of the degradation is worse
than that for the ASCA CCDs, despite that ASCA has been in a low-earth
orbit since February, 1993.  It should be noted that the FI CCDs are
heavily damaged while the BI CCDs are free from damage.  It was
concluded that the degradation was due to a relatively high flux of low
energy ($\sim100$ keV) protons. The interpretation is that the low
energy protons are collected onto the CCD and penetrate the relatively
thin electrode structure leaving traps near the charge transfer channel.

The charge transfer channel is a relatively narrow path inside the pixel
that is localized as a \lq notch\rq\ structure~\cite{burke97}.  The
effective thickness of the electrode structure above the charge transfer
channel is directly measured by the mesh experiment and found to be
$\sim0.3\,\mu$m thick Si and $\sim0.3\,\mu$m thick
SiO$_2$~\cite{pivovaroff98}.  Therefore, the high flux of low energy
protons penetrates this structure and causes permanent damage on the
buried channel of the CCD.  In the case of the BI CCD, the charge transfer
channel is relatively far away from the entrance side, resulting in no
damage to these CCDs.

The EPIC MOS CCD has a complicated electrode structure.  The effective
thickness of the absorber above the open electrode structure is designed to
be $\sim0.085\,\mu$m thick SiO$_2$.  This indicates that low energy
protons will easily generate traps under the open electrode structure.
However, the charge transfer channel is just under the finger structure.
Our measurements indicate that the finger structure consists of
$\sim0.2\,\mu$m thick Si and $\sim0.7\,\mu$m thick SiO$_2$.  This
indicates that the charge transfer channel of the EPIC MOS CCD is better
protected than that of the ACIS CCD.  We can say that the EPIC-MOS CCD
is more radiation hard than the FI chip of the ACIS.  The details on the
radiation hardness requires more quantitative measurement.


\subsection{Application}
X-ray events detected by a CCD are classified by the number of pixels     
over which the resultant charge splits.  Due to readout noise
and charge loss around the charge cloud perimeter, single pixel events
generally give the best energy resolution.  One of the main features of
this CCD is its open electrode region where the low energy efficiency is
enhanced.  We must therefore endeavour to control the operating conditions
so that we can increase the active region generating single pixel events
in the open electrode.

\begin{figure}[htbn]
\begin{center}
\vspace{-30mm}
\hspace{-7.3mm}
\psbox[r,xsize=0.48#1]{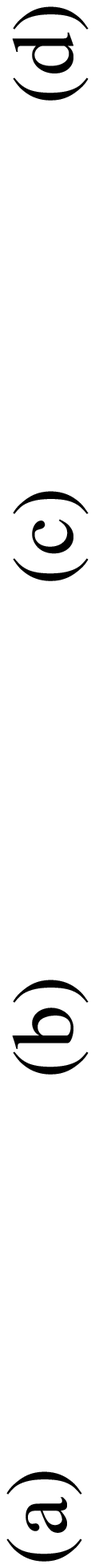}
\vspace{-90mm}
\psbox[r,xsize=0.48#1]{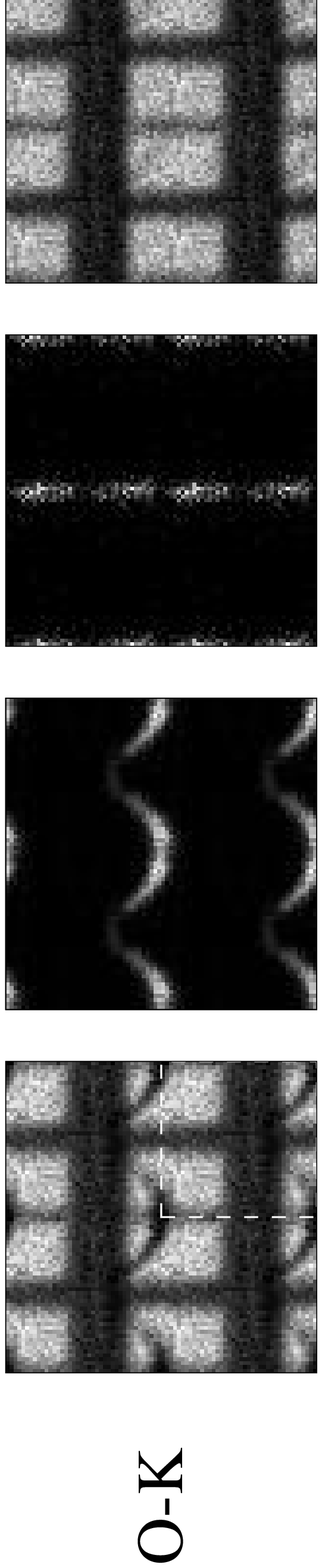}
\vspace{-68mm}
\psbox[r,xsize=0.48#1]{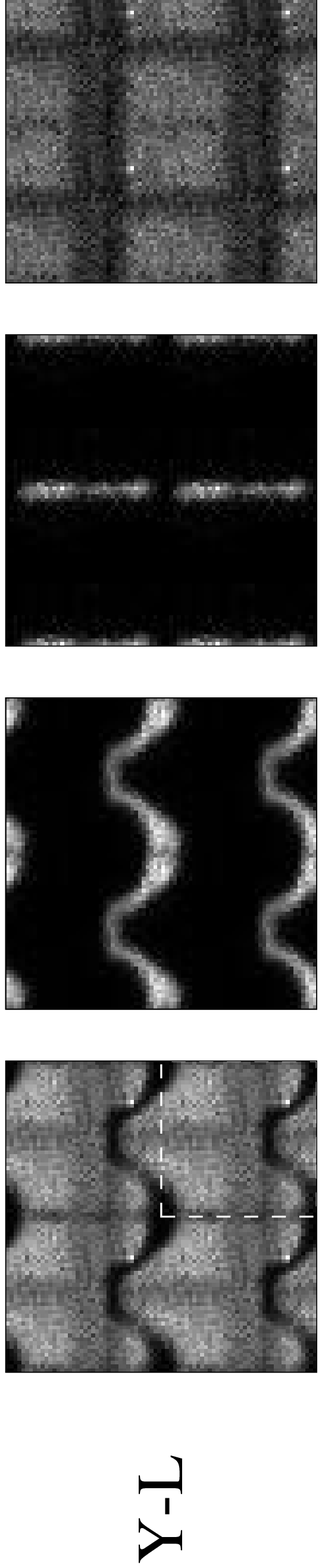}
\vspace{-68mm}
\psbox[r,xsize=0.48#1]{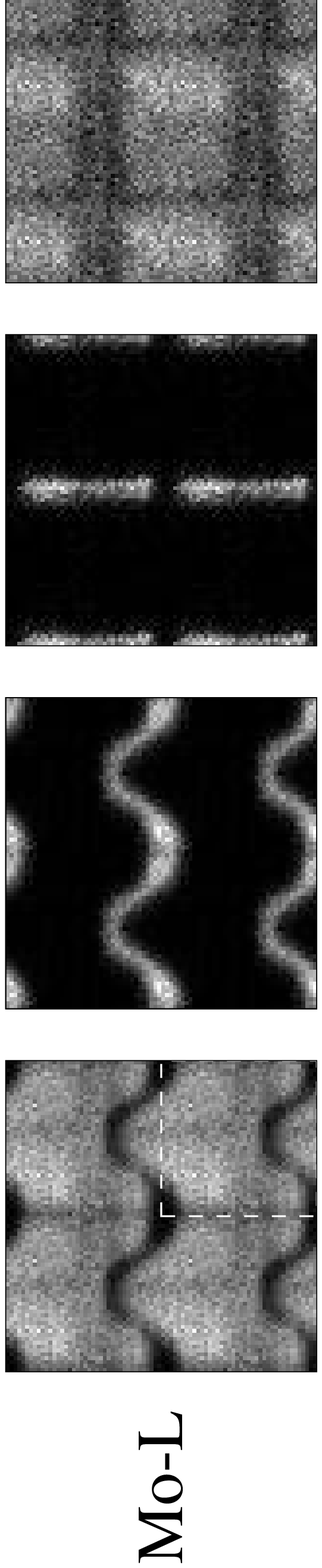}
\vspace{-68mm}
\psbox[r,xsize=0.48#1]{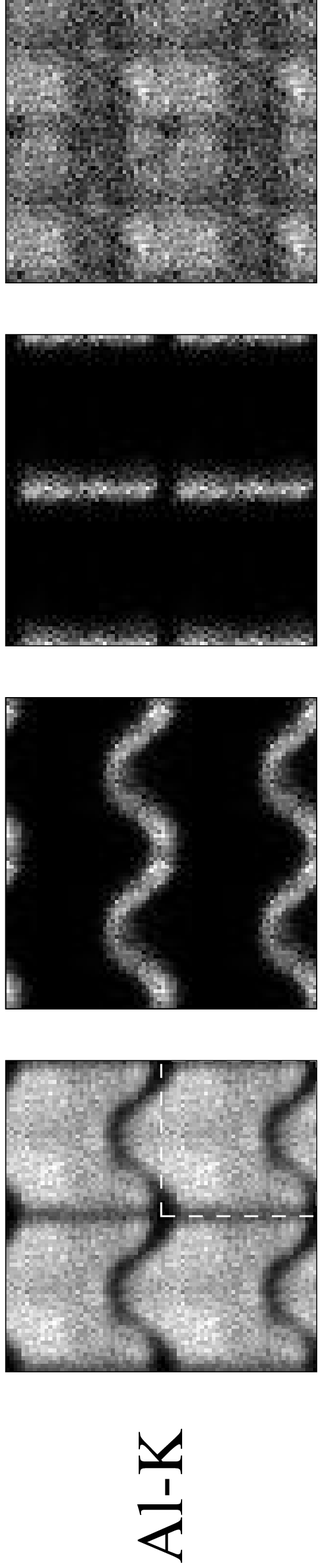}
\vspace{-68mm}
\psbox[r,xsize=0.48#1]{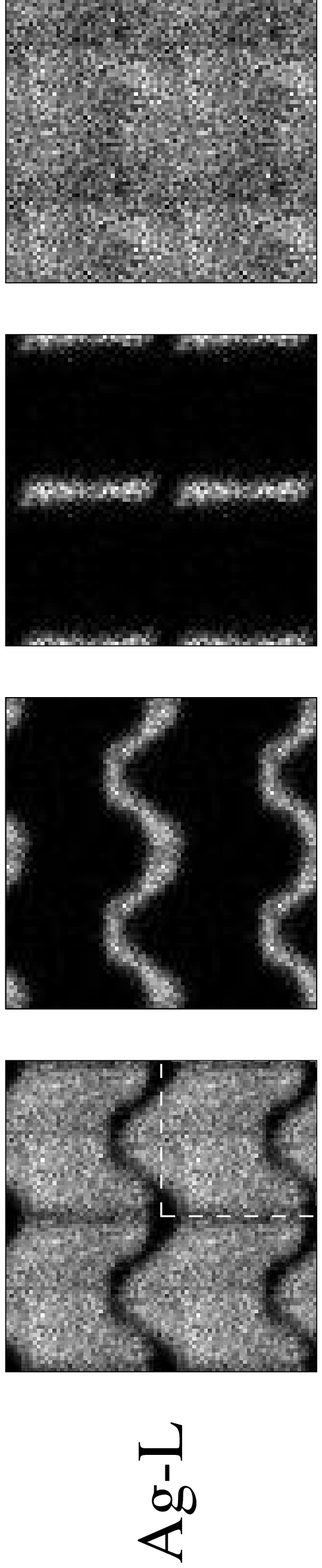}
\vspace{-40mm}
\caption{Restored images using characteristic X-ray emission lines for
 various X-ray event-types; (a) single pixel events, (b) vertically
 split events, (c) horizontally split events and (d) all X-ray
 events. The results are sorted by the order of the attenuation length
 in SiO$_2$. Each image represents 2$\times$2 pixels of the CCD, with
 the dashed square corresponding to the pixel size of 40\,$\mu$m square. 
 } 
\label{restored} 
\end{center}
\end{figure}

\begin{figure}[htbn]
\begin{center}
\psbox[r,xsize=0.48#1]{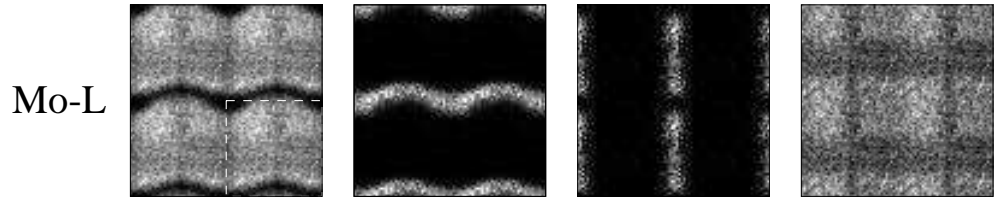}
\caption{Same as for Fig.4 for Mo-L (2.3\,keV) using the
 data with another cloking pattern during integration time.}
\label{rest2} 
\end{center}
\end{figure}

Fig.~\ref{restored} clearly shows that vertically split events are
generated primarily in the open electrode.  Furthermore, the shape of this
region depends on the attenuation length of the photon in Si.  This must
be due to the interaction of the applied voltages on neighbouring
electrodes.  In the standard mode, poly3 is biased during integration
while the other two electrodes are not.  We performed an experiment in which
we biased both poly3 and poly1 during integration. Figure~\ref{rest2} shows
the restored images for Mo-L(2.3keV) X-ray events again classified by
X-ray event-type as in Fig.~\ref{restored}.  It is clear that the
horizontal pixel boundary for vertically split events has been
changed.  We have not yet been able to determine which working condition
is the best for spectroscopic study. This will require further study,
taking into account the effect of the thickness of the depletion region.

In the mesh experiment, we can study how the primary charge behaves inside
the CCD~\cite{mark99}.  In some pixel regions, when the primary charge is
photoabsorbed close to the front surface of the device, it produces a tail
in the spectrum.  This effect is more evident for low energy X-rays since
they have a shorter attenuation length in Si. In particular, the response
of the O-K line shows a strong dependence on the interaction position
within the pixel.  In our experiment, the incident X-ray spectrum contains
several emission lines as well as a relatively strong continuum which
prevents us from studying in detail any tail to the response function. In
order to study this, we require a mono-energetic X-ray beam. The present
experimental setup does not currently permit us to generate such mono-energetic
X-ray beams with sufficient flux to conduct the mesh experiment.

\subsection{Detection Efficiency}
\begin{figure}[htbn]
\begin{center}
\psbox[r,xsize=0.35#1]{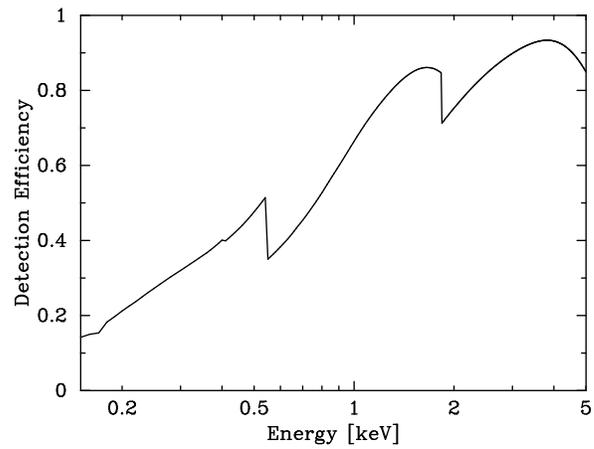}
\caption{Mean pixel response calcurated based
on our results.}
\label{eff} 
\end{center}
\end{figure}

We calculated the meadn pixel response that is shown in Figure~\ref{eff}
 with taking into acount our results of each structre within the
 pixel;\,open-electrodes, electrodes, finger and channel stop. It was
 calcurated with parameters measured from the mesh experiment as
 described in table~\ref{result} weighted with area fractions of each
 structure.  In this calculation, we assumed that there is an extra
 absorption of $\sim0.085\,\mu$m thick SiO$_2$ and Si$_3$N$_4$ over the
 pixel since we can not measure the absorption on the open electrode
 region.  These extra absorption play an important role of the
 absorption feature at the low energy region.  We also assume that the
 depletion region is 37$\mu$m~\cite{alex98} thick that plays an
 important role at the high energy region.

We compared our result with that reported by C.~Pigiot et
al. 1999~\cite{pigiot99} which performed the ground caribrations of EPIC
MOS CCD using mono-energetic X-ray beam at Osay synchrotron facility in
France.  We found that our estimation has a good agreement with the data
points of thier ground calibration.

\section{Conclusion}
We performed a mesh experiment on a CCD CCD identical to that employed
in the EPIC MOS imaging spectrometers on-board XMM/NEWTON.  We were able
to obtain restored X-ray images with sub-pixel resolution using X-ray
photons of characteristic emission lines: O-K, Y-L, Mo-L, Al-K and Ag-L.
All the X-ray events are classified by their event-types: single pixel
event, horizontally split event, vertically split event and all events.
There are clear absorption features inside the pixel including electrodes,
channel stops etc. The shorter the attenuation length in SiO$_2$,
the clearer these absorption features become.

We also confirmed that the horizontal pixel boundary between vertically
split events is not a straight line, but is \lq wavy\rq.  The specific
shape of this wavy line depends on the attenuation length of the X-ray
photons in silicon. This indicates the depth dependence of the electric
field inside the CCD.  We obtained spectra from various regions within a
CCD pixel, which showed a non-uniformity of the detection efficiency.
We selected five regions from which we extracted spectra.  We then
compared the spectrum from the open electrode region with the other
regions in order to measure the absorption features in detail.  The
electrode structure comprises 0.29 $\pm 0.03\, \mu$m of Si and 0.94 $\pm
0.05\,\mu$m of SiO$_2$ while the electrode finger comprises 0.15 $\pm
0.05\,\mu$m of Si and 0.73 $\pm 0.02\,\mu$m of SiO$_2$.  The effective
absorption at the channel stop is equivalent to 0.57 $\pm 0.03\,\mu$m of
SiO$_2$.

The charge transfer channel is a relatively narrow path within the
pixel;just below the electrode finger in the case of EPIC MOS CCD. We
found that the extra absorption feature above the charge transfer
channel of EPIC MOS CCD is thicker than that of the FI CCD of
ACIS($\sim$0.3$\mu$m of Si and SiO$_2$) on-board the CXO whose orbit is
similar to that of XMM/NEWTON.  For the ACIS CCD, it was reported that
FI CCDs got permanent damages on the buried channel due to a
relatively high flux of low energy protons. Our measurement indicates
that the EPIC MOS CCD is more radiation hard than the FI CCD of the ACIS.

In the standard working condition of the CCD, the vertically split
events are generated mainly in the open electrode region where the
detection efficiency at low energies is enhanced.  Since single pixel
events usually give better energy resolution than split pixel events, it
is preferable to generate more single events and less split events in
the open electrode region.  We confirmed that the shape of the region
generating the vertically split event varies by changing the operation
mode. We can therefore control the CCD so that the open electrode region
generates more single pixel events rather than split events.

We calcurated the mean pixel response based on our measurement. Our
estimation has a good agreement with that performed on the ground
calibration of the EPIC MOS CCD using mono-energetic X-ray beam.

\ack
 This research was partially suported by Simitomo Foundation.

\newpage
\begin{table}
\vspace*{7cm}
\caption{Summary of characteristic X-ray energies obtained.}
\label{eev_exp}
\begin{tabular}{cccc} \hline
Characteristic X-ray & Energy [keV] & Attenuation  & Attenuation   \\
& & length in Si [$\mu $m] & length in $\rm{SiO_2}$[$\mu$m] \\
\hline
O-K & 0.52 & 0.47 & 0.93 \\
Y-L & 1.9 & 1.4 & 2.4 \\
Mo-L & 2.3 & 2.2 & 3.9 \\
Al-K & 1.5 & 7.9 & 4.2 \\
Ag-L & 2.9 & 4.4 & 7.9 \\
\hline
\end{tabular}
\end{table}

\newpage
\begin{table}[n]
\vspace*{7cm}
\caption{The extra thickness of various regions within the pixel}
\label{result}
\begin{tabular}{ccc} \hline
Selected region & thickness of Si [$\mu $m] & thickness in $\rm{SiO_2}$[$\mu$m] \\
\hline
Electrodes  & 0.29 $\pm 0.03 $ & 0.94 $\pm 0.05 $ \\
Finger & 0.15 $\pm 0.05$ & 0.73 $\pm 0.02 $ \\
Channel stop & $-$  & 0.57 $\pm 0.03 $ \\
\hline
\end{tabular}
\end{table}

\newpage

\end{document}